\documentclass{iopart}

\usepackage{ifthen}
\usepackage{ifpdf}
\usepackage{color}

\usepackage{latexsym}
\usepackage{amstext}
\usepackage{amssymb}
\usepackage{bm}

\ifpdf
\usepackage{graphicx}
\usepackage{epstopdf}
\else
\usepackage{graphicx}
\usepackage{epsfig}
\fi

\newcommand{\rmrk}[1]{#1}
\newcommand{\Eq}[1]{\textcolor{blue}{Eq.~(\ref{#1})}} 
\newcommand{\Fig}[1]{\textcolor{blue}{Fig.}~\ref{#1}}
\newcommand{\Cite}[1]{\textcolor{blue}{\cite{#1}}}

\newcommand{\im}{\mbox{Im}}

\newcommand{\eexp}{\mbox{e}^}

\newcommand{\mass}{\mathsf{m}}

\newcommand{\mylabel}[1]{\label{#1}} 
\newcommand{\beq}{\begin{eqnarray}}
\newcommand{\eeq}{\end{eqnarray}} 
\newcommand{\be}[1]{\begin{eqnarray}\ifthenelse{#1=-1}{\nonumber}{\ifthenelse{#1=0}{}{\mylabel{e#1}}}}
\newcommand{\ee}{\end{eqnarray}} 

\newcommand{\hide}[1]{}

\newcommand{\mpg}[2][1.0\hsize]{\begin{minipage}[b]{#1}{#2}\end{minipage}}

\begin{document}

\title[Quantum transport]{Multiple path transport in quantum networks}

\author{Geva Arwas, Doron Cohen}

\address{\mbox{Department of Physics, Ben-Gurion University of the Negev, Beer-Sheva 84105, Israel}}

\begin{abstract}
\rmrk{We find an exact expression for the current~($I$) 
that flows via a tagged bond from a site (``dot") 
whose potential~($u$) is varied in time.}
We show that the analysis  
reduces to that of calculating time dependent probabilities, 
as in the stochastic formulation, but with splitting (branching) 
ratios that are not bounded within ${[0,1]}$.
\rmrk{Accordingly our result can be regarded 
as a multiple-path version of the continuity equation.
It generalizes results that have been obtained 
from adiabatic transport theory in the context 
of quantum ``pumping" and ``stirring".} 
Our approach allows 
to address the adiabatic regime, 
as well as the Slow and Fast 
non-adiabatic regimes, on equal footing. 
We emphasize aspects 
that go beyond the familiar picture of sequential Landau-Zener crossings, 
taking into account Wigner-type mixing of the energy levels. 
\end{abstract}

\section{Introduction}

Transport in quantum networks is a theme that emerges in diverse contexts, 
including quantum Hall effect \Cite{hall}, Josephson arrays \Cite{JJ},  
quantum computation models \Cite{q_compute}, 
quantum internet \Cite{q_internet},   
and even in connection with photosynthesis \Cite{plants}. 
For some specific models there are calculations 
of the induced currents in the adiabatic regime \Cite{Th1,Th2,Berry,Avron,Robbins}
for both open and closed systems, 
so called ``quantum pumping" \Cite{BPT,pmp1,pmp2,pmp3,pmp4,pmo,pmp5,pmp6} 
and ``quantum stirring" \Cite{pmc,pmt,pml,pms,pmr} respectively. 
In the latter context most publications focus on 2-level \Cite{zener1,zener2}
and 3-level dynamics, while the larger perspective is 
rather abstract, notably the ``Dirac monopoles picture" \Cite{Avron,pmc,pml,pms}.
This should be contrasted with the analysis of 
stochastic stirring where the theory is quite mature \Cite{Saar,st1,st2,st3}.

\begin{figure}
\centering

\mpg[7cm]{
\includegraphics[width=5cm]{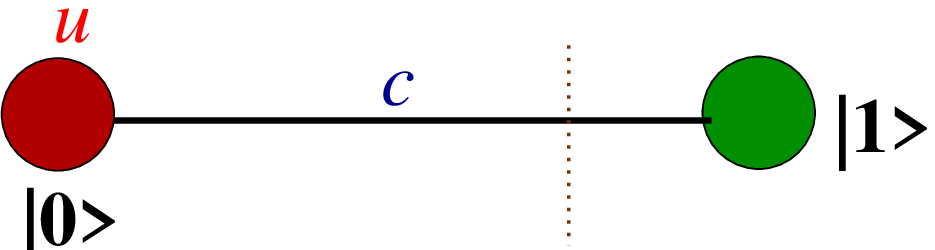} 

\ \\

\includegraphics[width=5cm]{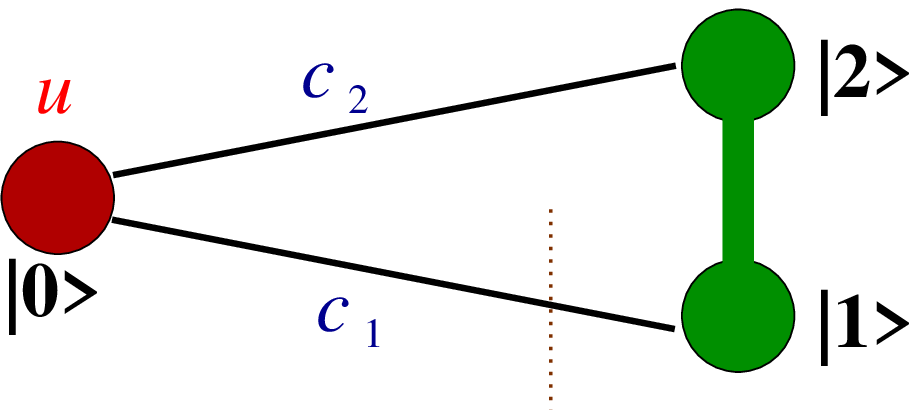} 
}
\includegraphics[width=5cm]{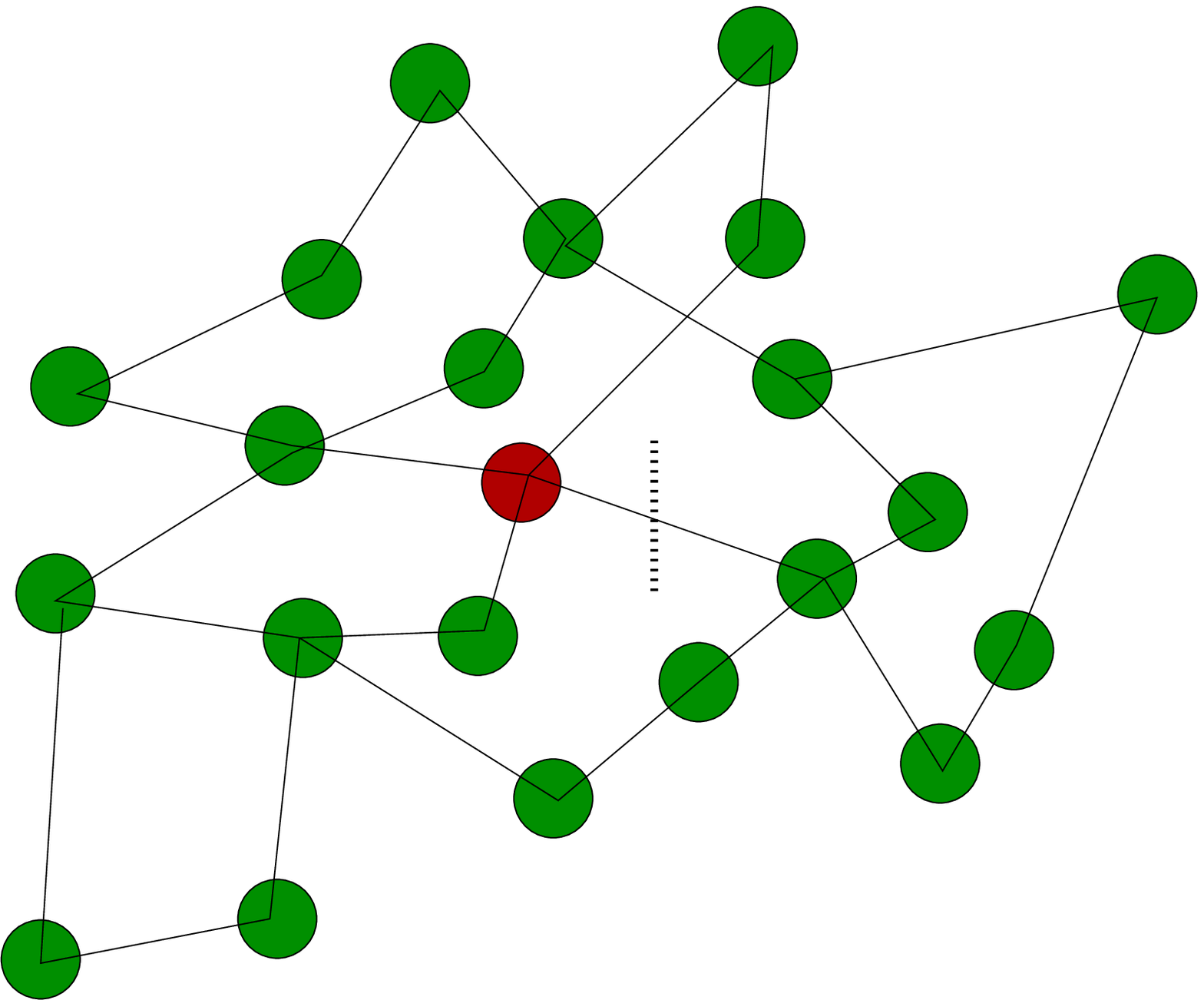} 

\caption {
Two-site, three-site, and $1{+}N$ site models for transport. 
A particle is initially positioned at the ``dot" site $|0\rangle$, 
The dot has a potential energy~$u$ that can be controlled externally. 
As~$u$ is varied, say from $-\infty$ to $+\infty$, 
currents are induced in the bonds of the network. 
Our objective is to calculate the current that flows 
from the dot to the network via a tagged bond. 
The vertical dotted line indicates a section through 
which the current of interest is flowing. 
}

\label{fModels}
\end{figure}


\begin{figure}

\includegraphics[width=15cm,clip]{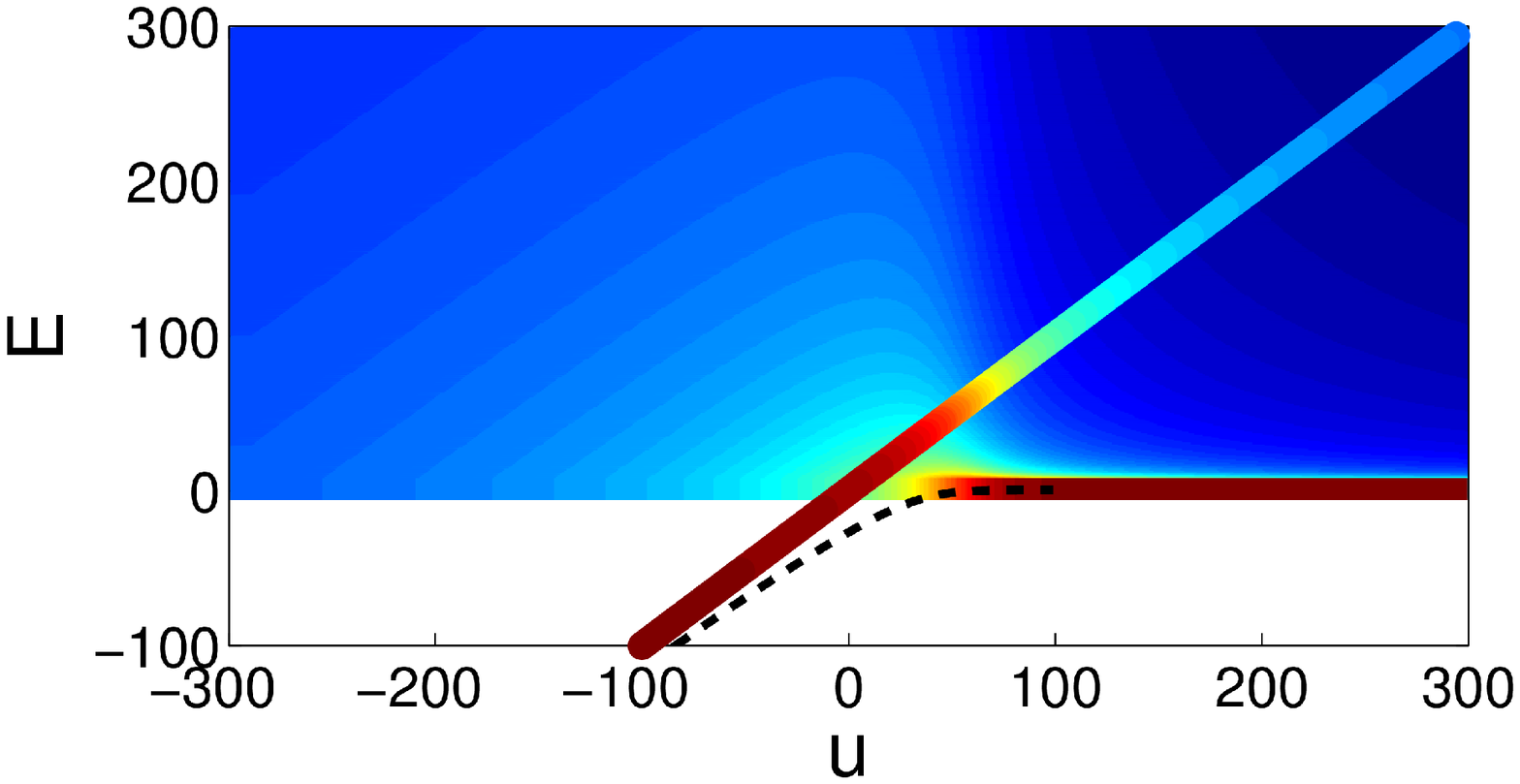}
\centering

\setlength{\unitlength}{1.5mm}
\begin{picture}(60,60)
\put(0,0){\includegraphics[width=90mm,clip]{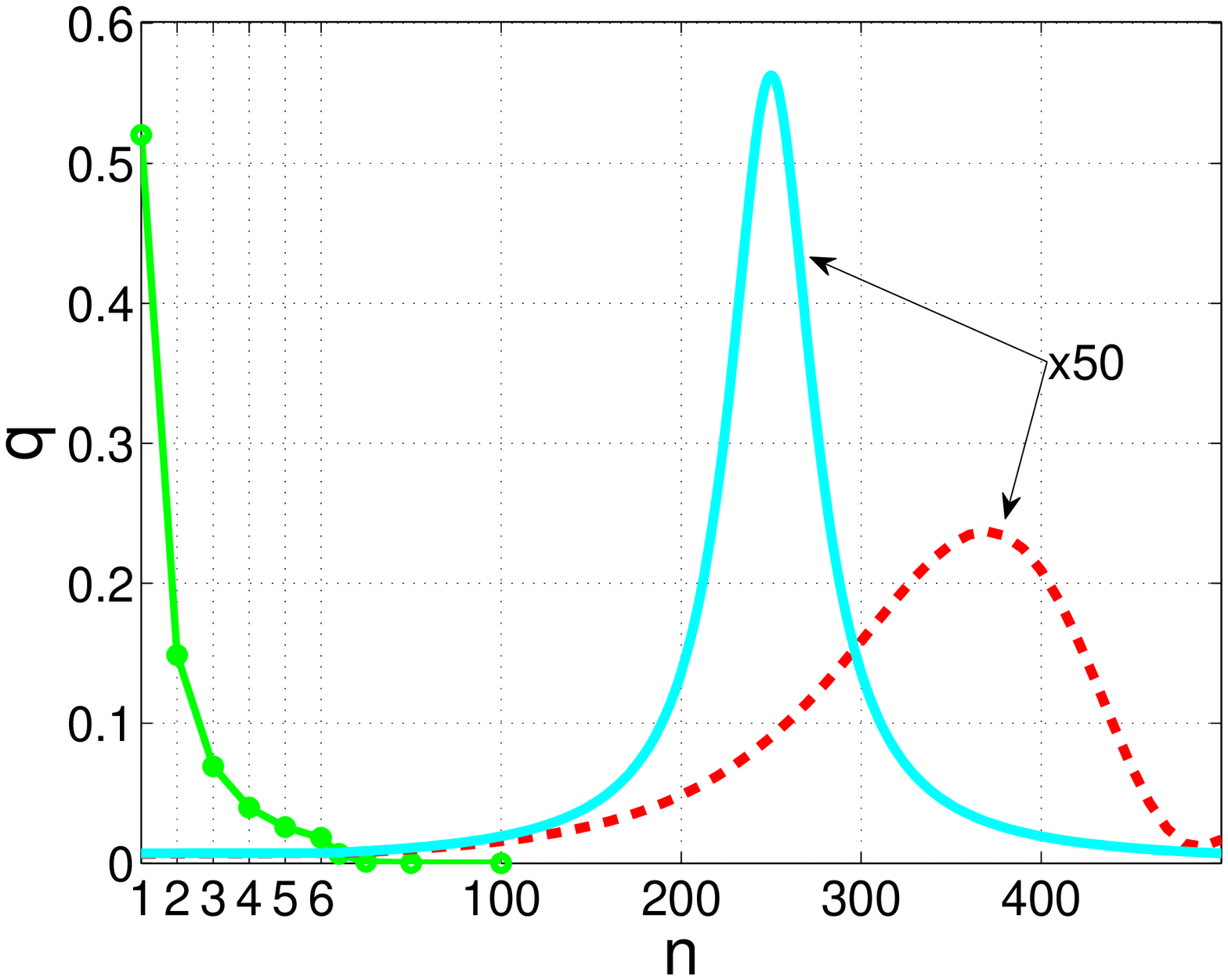}}
\put(15,25){\includegraphics[width=22.5mm,clip]{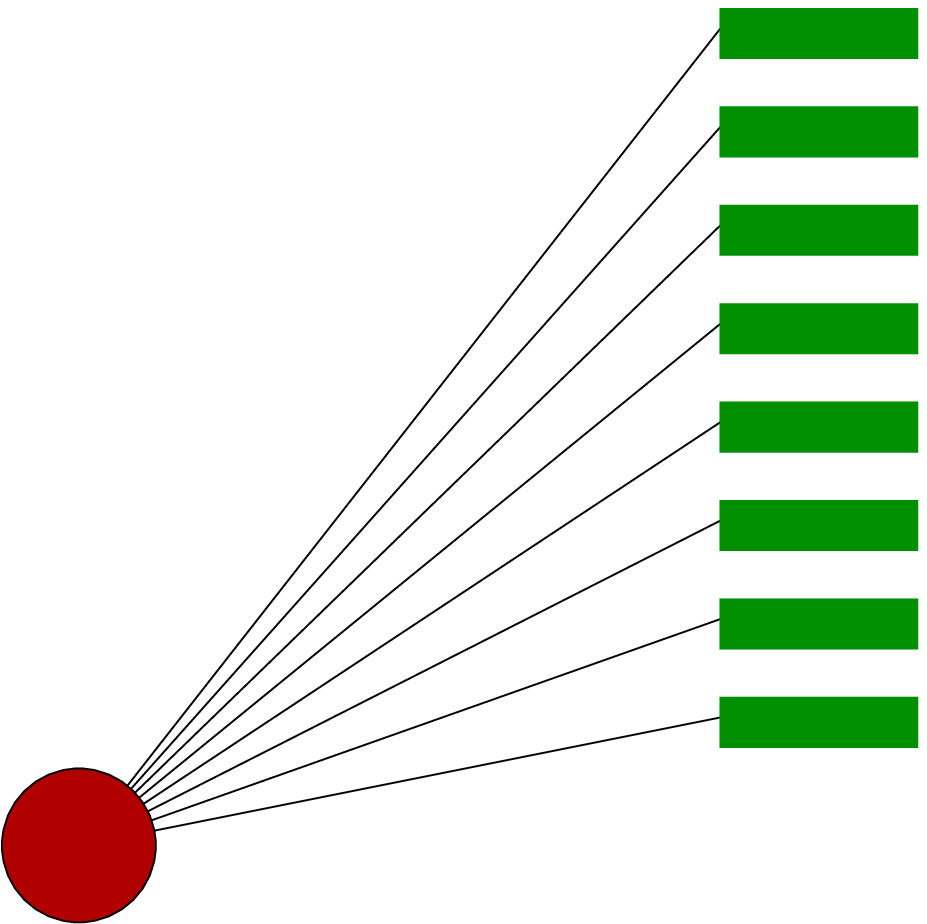}}
\end{picture}

\caption{
{\em Upper panel:} 
an initially loaded ``dot" level is crossing 
a band that contains $N=500$ network levels.   
The occupation probabilities of the dot ($p$) 
and of the levels ($q_n$) are imaged as a function 
of time.
The vertical axis is the energy,  
and the horizontal axis is $u(t)$.
\rmrk{In this specific example 
the sweep process is adiabatically slow:  
As $u$ is increased the dot level gets 
emptied (color changing from red to blue), 
while at the end of the process 
only the ground level of the network is 
occupied (color changing from blue to red).}
We assume star geometry with level spacing $\Delta{=}1$, 
and identical couplings ${c_n{=}3}$.
The dashed line illustrates the energy of the lowest adiabatic level.
{\em Lower panel:} 
plot of $q_n$ vs $n$ in several cases. 
Green line with markers - the adiabatic scenario of the upper panel at $u=60$.
Cyan solid line - after decay from a standing level ($\dot{u}{=}0$).
Red dashed line - after decay from a moving level ($\dot{u}{=}5000$). 
}

\label{fS}
\end{figure}


In this work we would like to analyze the following prototype 
problem. Consider a network \rmrk{as illustrated in \Fig{fModels}.}
It consists of $N$ interconnected sites, 
with on-site energies $\mathcal{E}_i$, and couplings $C_{ij}$.
Additionally there is a site ($i=0$) \rmrk{that we call ``dot"}, 
where the potential energy $\mathcal{E}_0=u(t)$ 
is varied \rmrk{according to some time-dependent protocol. 
For illustration purpose we assume that the on-site potential 
is {\em swept} monotonically from $u=-\infty$ to $u=\infty$.} 
The Hamiltonian is 
\beq  
\mathcal{H} \ \ = \ \ \sum_{i=0}^N |i\rangle \mathcal{E}_i \langle i| 
\ + \ \sum_{i \ne j} |i\rangle C_{ij} \langle j |   
\\
\ \ \ \ \ \mathcal{E}_0{=}u(t), \ \ C_{i0}{=}C_i
\eeq
Our interest is in the induced current $I(t)$ 
that flows through a tagged bond ${0\leadsto a}$
\rmrk{that connects the dot (${i=0}$) with some other network site (${i=a}$).} 
This bond is reflected in the Hamiltonian 
by the presence of a coupling constant $C_{a0}=C_a$.

\rmrk{In order to have a well-posed problem we assume that there 
is no magnetic field: accordingly all the couplings can be gauged 
as real numbers; and there are no persistent currents in the network.} 
In the adiabatic limit \Cite{Th1,Th2,Berry,Avron,Robbins,pmc,pmt,pml,pms,pmr}   
the current $I(t)$ is proportional at any moment to $\dot{u}$, 
and can be calculated as follows: 
\be{1}
I=G\dot{u}, 
\hspace{15mm}
G =  
2\im \left[ \Big\langle 
\frac{\partial}{\partial \phi} \Psi \Big| \frac{\partial}{\partial u} \Psi 
\Big\rangle \right]_{\phi=0}
\eeq
Here $\phi$ is a {\em test} flux through the bond of interest, 
namely $C_a\mapsto C_a \eexp{i\phi}$, and $\Psi$ is the 
wave-function of the adiabatic eigenstate. 
The coefficient $G$ is known as the {\em Geometric Conductance}, 
or as the {\em Berry-Kubo curvature}.  
In \Cite{pmr} the interested reader can find how this formula 
is used in order to determine the current in 
the two-site and three-site models that are illustrated in \Fig{fModels}.

\rmrk{The adiabatic transport formula \Eq{e1} is not transparent: 
it requires some effort to get a heuristic understanding of its outcome. 
Furthermore it does not apply to non-adiabatic circumstances.
We therefore look for a different way of calculation.  
Evidently for a two-site model, as illustrated in \Fig{fModels}, 
we can simply use the continuity equation:}
\beq
I = 
\frac{\partial}{\partial t} \left[ q_1 \right],
\hspace{15mm}
q_1 = |\langle 1|\Psi(t)\rangle|^2 
\eeq
\rmrk{where $q_1$ is the occupation probability of the $i=1$ site.
Clearly, this formula holds irrespective of whether 
the sweep process is adiabatic or not. Hence the problem 
of calculating currents trivially reduces to the calculation  
of a time-dependent occupation probability. }

\rmrk{Considering a general network, our main observation is 
that for a multiple-path geometry the continuity equation 
can be generalized as follows:}
\be{6}
I =
\frac{\partial}{\partial t}
\left[ \sum_n \lambda_n   q_n \right],
\hspace{15mm}
q_n = |\langle \epsilon_n|\Psi(t)\rangle|^2 
\eeq
\rmrk{Here the $q_n$ are the occupation probabilities of the network levels $|\epsilon_n \rangle$, 
and the pre-factors $\lambda_n$ are determined by the coupling constants. 
We refer to $\lambda_n$ as the splitting ratio: it describes 
the relative contribution of the ${0\leadsto n}$ flow  
to the current in the tagged bond. Hence, again, the 
calculation of the current reduces to that of calculating 
time-dependent probabilities, as in the stochastic formulation. 
But we shall see that the splitting ratios, unlike the branching
ratios of the stochastic theory, are not bounded within ${[0,1]}$. }
For a non-interacting many-body occupation, 
results can be obtained by simple summation,  
with~$q_n(t)$ that represent the actual occupations of the levels. 

\rmrk{As already stated, for demonstration purpose, 
we are going to analyze a {\em sweep} process, 
in which the on-site potential is varied monotonically 
from $u=-\infty$ to $u=\infty$. We are going to 
distinguish between two {\em sweep scenarios}:} 
\begin{itemize}
\setlength{\itemsep}{0pt}
\item
{\em Injection -} the dot is initially filled 
with a particle that later is transferred to the network; 
\item
{\em Induction -} one of the levels of the network 
is initially filled,  and later a current is induced via the crossing dot.
\end{itemize}
The occupation dynamics in the first (Injection) scenario is illustrated 
in \Fig{fS}, which will be further discussed later.   
In later sections we consider also the second (Induction) scenario 
considering ``star geometry" and ``ring geometry".

\section{Star geometry, adiabatic limit}

Let us consider the special geometry of a network 
that consists of sites $\mathcal{E}_n=\epsilon_n$,  
and connections $C_{n0}=c_n$,
while all the other couplings are zero, 
as illustrated in the inset of \Fig{fS}. 
An adiabatic eigenstate $|\Psi\rangle$ is represented 
by a column vector ${\psi_n=\langle n|\Psi\rangle}$  
that satisfies the following set of equations:
\be{160}
u\psi_0+ \sum_{n=1}^N c_n^* \psi_n  &=& E\psi_0 
\\ \label{e170}
c_n \psi_0 + \epsilon_n \psi_n  &=&  E \psi_n, \ \ \ \ \ \ n=1,2,...,N 
\eeq 
It follows from \Eq{e170} that it can be written as:
\be{101} 
|\Psi\rangle    
\ \ = \ \ 
\sqrt{p} \, |0\rangle 
\  + \ 
\sqrt{p} \sum_{n=1}^N \frac{c_n}{E-\epsilon_n} \, |n\rangle 
\eeq
where $\sqrt{p}$ is a normalization constant.
We define  
\be{40}
g(E;c_1,...c_N) \ \ = \ \ \sum \frac{|c_n|^2}{E-\epsilon_n}
\eeq
Substitution of the $\psi_n$ of \Eq{e101} into  \Eq{e160}
leads to the secular equation ${g(E)=E-u}$
for the adiabatic eigen-energies. 
We focus our attention on a particular root $E(u)$.  
As $u$ is swept from $-\infty$ to $+\infty$, 
the energy $E(u)$ increases monotonically 
from $\epsilon_{n_0}$ to $\epsilon_{n_0+1}$, 
where $n_0$ is the starting level. 
From \Eq{e101} it follows that $p$ is 
the probability to find the particle in the dot. 
It can be written as 
\be{100} 
p(u) \ \ = \ \  |\psi_0|^2 \ \ = \ \ \Big[1-g'(E(u))\Big]^{-1}
\eeq
For the following derivation note that $1/p$ 
is a quadratic form in~$c_n$, and that 
the occupation probabilities of the 
network levels ${n=1,2,3...}$ are 
\be{102} 
q_n(u) \ \ = \ \ |\psi_n|^2 \ \ = \ \ \left|\frac{c_n}{E(u)-\epsilon_n}\right|^2 \  p(u) 
\eeq
Using \Eq{e1} we get after 
differentiation by parts that the 
current through $c_n$ is 
\beq \nonumber
G \ &=& \ 
\frac{|c_n|^2}{(E-\epsilon_n)^2}\left(\frac{\partial p}{\partial u}\right)
-2p\frac{|c_n|^2}{(E-\epsilon_n)^3}\left(\frac{\partial E}{\partial u}\right)
\\
\label{e4}
\ &=& \
\frac{\partial}{\partial u}
\left[
\left( \frac{1}{2}\frac{\partial (1/p)}{\partial c_n} c_n \right) \ p
\right]
\ \ = \ \
\frac{\partial}{\partial u}
\left[ q_n \right]
\eeq
We further discuss and generalize this trivial result below. 

\section{Multiple path geometry, adiabatic limit}

\begin{figure}
\centering

\includegraphics[height=3.5cm]{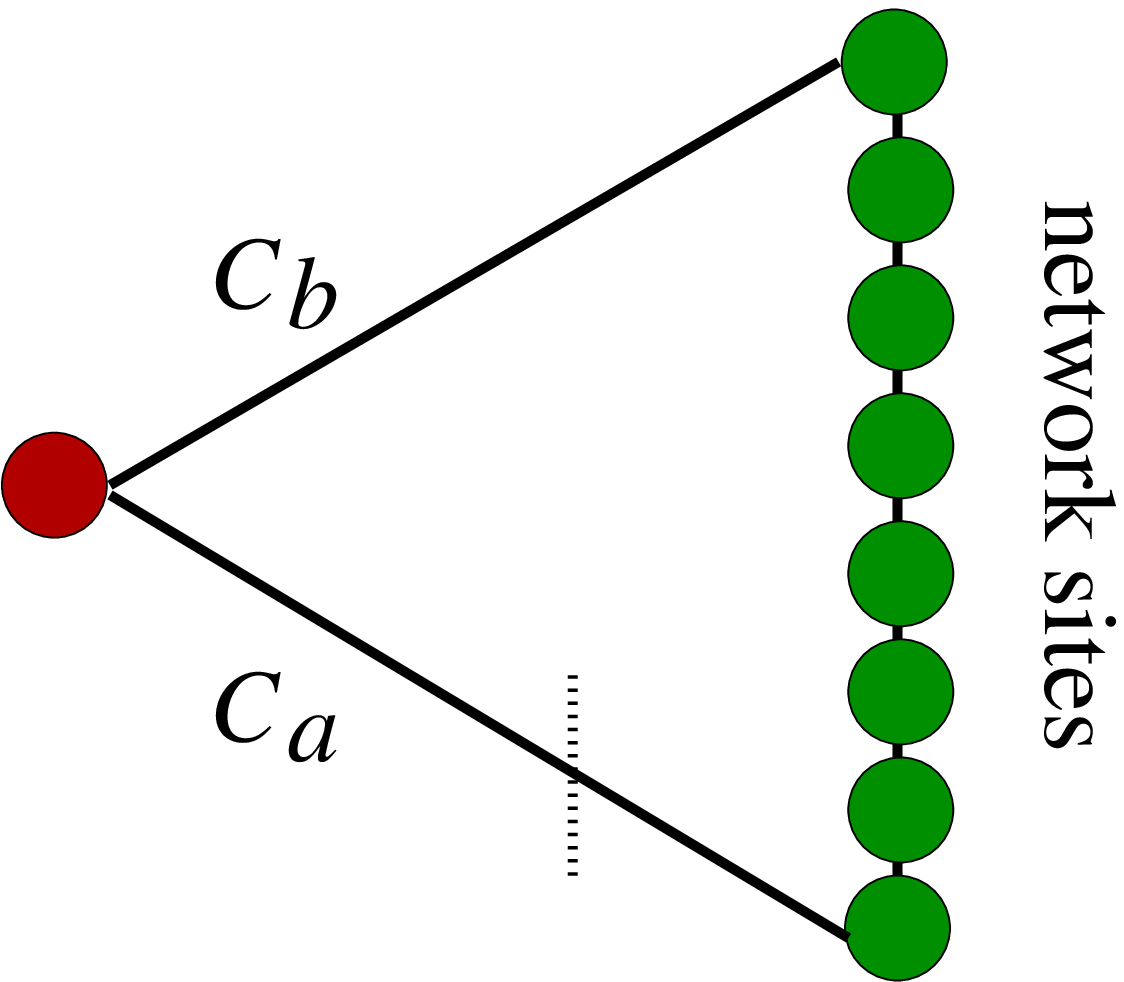} 
\ \ \ \ \ \ \ \ \
\includegraphics[height=3.5cm]{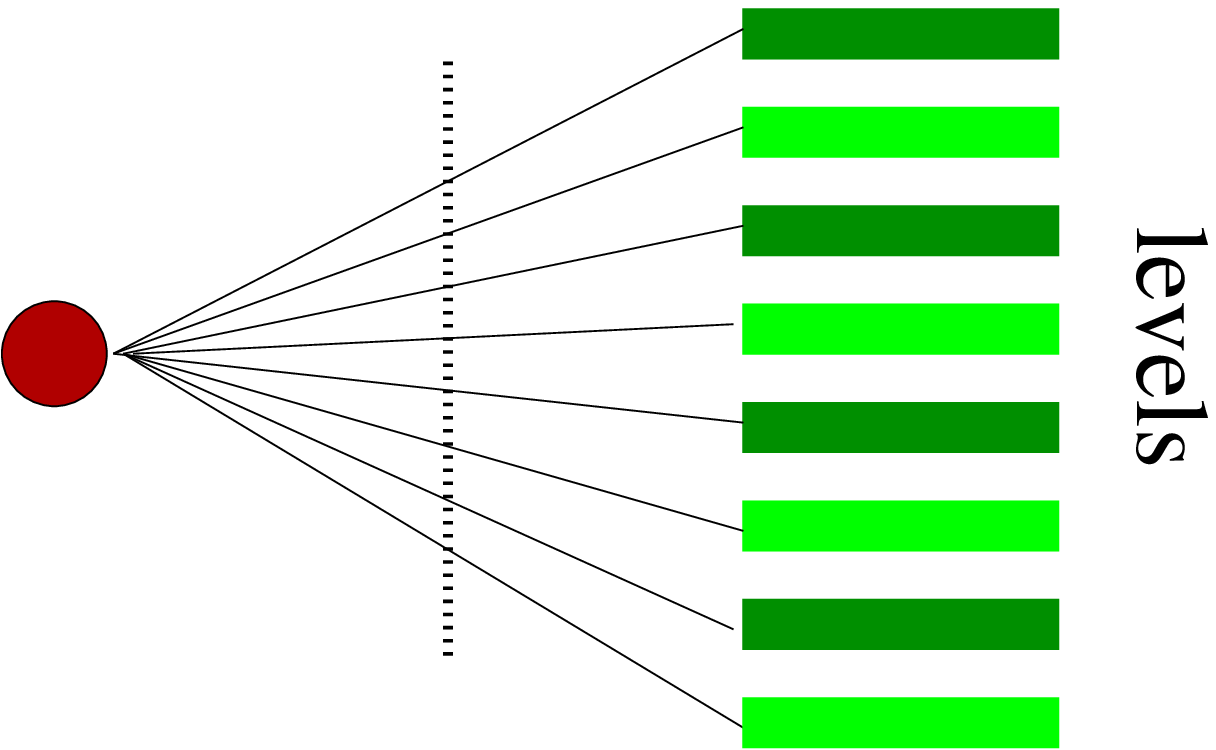} 

\caption {
The dot-wire ring geometry (left). If the Hamiltonian 
is written in the $n$ basis the problem reduces 
to star geometry. The couplings of the 
levels to the dot are ${c_n \propto (C_a \pm C_b)}$ 
for even or odd levels respectively. 
Hence it is like having {\em two} continua
rather than a single continuum. 
}

\label{fModelsDW}
\end{figure}

Let us find what is the expression for~$G$ in the case of 
a general network.  It is natural to switch from the $\mathcal{E}_i$ basis 
to an $\epsilon_n$ basis that diagonalize the network 
Hamiltonian in the absence of the dot.
Consequently getting a star geometry with 
\beq
c_n \ \ = \ \ \sum_{i} \langle \epsilon_n|i \rangle  C_{i}
\eeq
An example for this procedure is presented in Section~5 
with regard to the  dot-wire ring geometry of \Fig{fModelsDW}.
Our interest is in the current through a tagged bond $C_a$.
We define the ``splitting ratio" of the current 
that flows in the $n$th levels as  
\be{5}
\lambda_n[\text{splitting}] \ = \ \frac{\langle \epsilon_n|a \rangle  C_{a}}{c_n}     
\ = \ \frac{\langle \epsilon_n|a \rangle  C_{a}}{ \sum_{i=1}^N \langle \epsilon_n|i \rangle  C_{i} }
\eeq 
A straightforward generalization of the derivation
that leads to \Eq{e4} implies that the current 
through~$C_a$ is given by \Eq{e6}.

\rmrk{At this stage \Eq{e6} is regarded as the outcome 
of adiabatic transport theory, while in the next section 
we shall provide its general derivation, and observe 
that it is a valid result also in non-adiabatic circumstances.}

\section{Transport calculation - the splitting ratio approach}

Needless to say that we do not really need \Eq{e1} 
in order to get the expression for $G$ in the case 
of a star graph. We could simply deduce \Eq{e4} 
from  conservation of probability, 
i.e. from the continuity equation $I = \dot{q}_n$.
This is no longer the case if we have a multiple path 
geometry: probability conservation alone cannot tell us 
how the current is split between the different paths.
\rmrk{Inspecting \Eq{e6} it looks like a generalization 
of the continuity equation as discussed in the Introduction.}
Its physical simplicity suggests that it can be derived 
without assuming adiabaticity. 
\rmrk{We now show that this is indeed the case.}

\rmrk{The stating point is the assumption that we have in hand the 
solution of the  {\em time-dependent} Schrodinger equation, 
which can be written either in the $i$ or in the $\epsilon_n$ 
representations:  
\beq
|\Psi(t)\rangle    
\ \ &=& \ \ 
\Psi_0(t) \, |0\rangle + \sum_{i=1}^N \Psi_i(t) \, |i\rangle 
\\
\ \ &=& \ \
\psi_0(t) \, |0\rangle + \sum_{n=1}^N \psi_n(t) \, |\epsilon_n\rangle 
\eeq
We recall our definitions of occupation probabilities:
\beq
p(t) \ \ &=& \ \ \langle 0| \Psi(t)\rangle \ \ = \ \ \ |\Psi_0(t)|^2  \ \ = \ \ \mbox{dot occupation} \\
q_n(t) \ \ &=& \ \ \langle \epsilon_n| \Psi(t)\rangle \ \ = \ \ |\Psi_n(t)|^2 \ \ = \ \ \mbox{level occupations}
\eeq
and obviously the total occupation probability is unity:
\beq
p(t) \ + \ \sum_n q_n(t) \ \ = \ \ 1 
\eeq
In the $\epsilon_n$ basis the Hamiltonian becomes the same 
as in ``Star geometry". The current operator for the ${0\leadsto n}$ bond is  
\beq
\mathcal{I}_{0 \leadsto n} \ \ = \ \ -i c_n \, \Big[|n\rangle \langle 0| - |0\rangle \langle n|\Big]
\eeq
Accordingly we can write the continuity equation 
\be{201}
\dot{q}_n \ \ = \ \  \langle \Psi | \mathcal{I}_{0 \leadsto n} | \Psi \rangle  
\ \ = \ \  c_n \, \im\left[\psi_n^*\psi_0\right]
\eeq
But our interest is in the current that flows 
in real space through the tagged bond
\be{202}
I \ \ = \ \  \langle \Psi | \mathcal{I}_{0 \leadsto a} | \Psi \rangle 
\ \ = \ \ C_a \ \im\left[\Psi_a(t)^*\Psi_0(t)\right]
\eeq
The amplitudes $\Psi_i$ are related to the amplitudes $\psi_n$. In particular  
\be{203}
\Psi_a(t)= \sum_n \langle a|\epsilon_n \rangle \ \psi_n(t), 
\ \ \ \ \ \ \ \ \ \Psi_0(t)=\psi_0(t)
\eeq
Substitution of \Eq{e203} into \Eq{e202} gives
\beq
I \ \ = \ \  C_a  \ \im\left[\sum_n  \langle \epsilon_n |a \rangle \ \psi_n(t)^* \ \psi_0(t) \right]
\eeq
Using the identification of $\dot{q}_n$ from \Eq{e201} we get 
the desired result \Eq{e6} with \Eq{e5}. 
This very simple, and yet very general result, 
has far reaching consequences as described below.}


\section{The dot-wire ring geometry}

\rmrk{In order to demonstrate the application of the splitting ratio approach 
we shall consider the simplest non-trivial example,  
regarding the dot-wire ring geometry of \Fig{fModelsDW}. 
The ring consists of a ``dot" whose potential $u(t)$ can be varied in time, 
and a ``wire" that consists of ${i=1,...,N}$ sites
with ${\mathcal{E}_i=0}$ and near-neighbor couplings $C_{ij}=C_0$. 
Optionally an appropriate procedure allows to take 
the limit $N\rightarrow\infty$ keeping the length of the wire ($L \equiv (1{+}N)a$) 
and the mass of the particle (${\mass \propto 1/(C_0 a^2)}$) fixed. 
But the mathematics is more transparent with a tight binding model.} 
 
The energy levels of the wire are $\epsilon_n=-2C_0\cos(k_n)$, 
where the wavenumbers are ${k_n=(\pi/L)n}$, with  $L=N{+}1$.
The respective couplings to the dot are 
\be{225}
c_n \ \ = \ \ \left[\left(\frac{2}{L}\right)^{1/2}\sin(k_n)\right] \ (C_a\pm C_b)
\eeq
where the $\pm$ reflects the parity of the level.   
It follows that the splitting ratios are 
\beq
\lambda_n \ \ = \ \ \lambda_{\pm} \ \ = \ \ \frac{C_a}{C_a \pm C_b} 
\ \ \ \ \ \ \ \ \ \mbox{for level with even/odd parity}
\eeq

\rmrk{In a later section we consider $N\gg1$ wire, 
and focus on levels with wavenumber  ${k_n \sim k}$ 
and energy ${\epsilon_n \sim E=-2C_0\cos(k)}$,  
that are located away from the band edges.
In order to allow analytical treatment we assume that 
the density of states in the energy window of interest 
can be approximated as constant. Accordingly one can 
regard the level spacing $\Delta$ as a free parameter.} 
In the same spirit it is convenient to absorb 
the constant pre-factor in \Eq{e225}
into the definition of $C_a$ and $C_b$, 
such that $c_n= (C_a\pm C_b)/\sqrt{2}$.


\section{The integrated current}

\rmrk{From \Eq{e6} it follows that the integrated current 
after a sweep process can be calculated as follows:}
\beq
Q_{0 \,\leadsto\, a} 
\ \ \equiv \ \ \int I(t') \, dt' 
\ \ = \ \ \sum_{n} \Big[q_{n}(\text{final}) - q_{n}(\text{initial}) \Big] \ \lambda_{n} 
\eeq 
\rmrk{In particular for an Injection process}
\beq
Q_{0 \,\leadsto\, a} [\mbox{injection}]
\ \ = \ \ \sum_{n} q_{n}(\text{final}) \ \lambda_{n} 
\eeq 
For an {\em adiabatic} injection scenario, 
in which the particle ends up at the lower network level we get
\beq
Q_{0 \,\leadsto\, a} [\mbox{adiabatic injection}]
\ \ = \ \ \lambda_{\mbox{ground level}}
\eeq 
while in the non-adibatic case the sum can be 
regarded as a weighted average of the $\lambda_n$.
Let us consider for example the dot-wire ring system.
For an adiabatic injection scenario we get 
\beq
Q_{0 \,\leadsto\, a} [\mbox{adiabatic  injection}] \ \ = \ \ \lambda_{-} \ \ = \ \ \frac{C_a}{C_a{-}C_b}
\eeq
Unlike the case of a stochastic transition 
this value is not bounded within ${[0,1]}$. 
rather it may have any value, depending on the 
relative sign of the amplitudes $C_a$ and $C_b$. 
But if the process is not adiabatic, the probability is 
distributed over both the odd and the even levels 
with probabilities that are proportional 
to ${|C_a \pm C_b|^2}$ respectively. 
Then we get from the weighted average a stochastic-like
result, namely  
\beq
Q_{0 \,\leadsto\, a}[\mbox{fast injection}] \ \ = \ \  \text{average}(\lambda_n)  \ \ = \ \  \frac{|C_a|^2}{|C_a|^2+|C_b|^2}
\eeq

For an adiabatic {\em induction} scenario,
the particle is prepared (say) in an even wire-level, 
and is adiabatically transferred, due to the sweep, 
into the adjacent odd wire-level. Then we get  
\beq
Q_{0 \,\leadsto\, a}[\mbox{adiabatic induction}]  \ \ = \ \  \lambda_{-}-\lambda_{+}  \ \ = \ \  \frac{2C_aC_b}{|C_a|^2-|C_b|^2}
\eeq
It looks as if the result does not depend on~$C_0$.
However this is misleading. In the next section 
we shall give a detailed account with regard 
to the time dependence of~$I$, and we shall see 
that the induction process is significantly different 
depending on whether $C_0$ is small or large.


\section{The parametric variation of the current}

The results for the integrated current give the impression 
that the size of the coupling $c_n$ compared 
with the levels spacing $\Delta$ is of no importance. 
But this is a wrong impression. Once we get deeper 
into the analysis it becomes clear that the familiar 
two level approximation for the adiabatic current~$I$, 
requires the coupling $c_n$ to be very small 
compared with the level spacing $\Delta$.
Our interest below is focused in the case 
of having a quasi-continuum, meaning 
that the $c_n$ are larger than $\Delta$, 
hence many levels are mixed \rmrk{during the sweep process}. 

Before discussing the quasi-continuum case it is useful 
to note that the 3~site ($N=2$) ring system has been  
solved exactly in \Cite{pmr}. It has been found that 
if the $c_n$ are not smaller compared with $\Delta$, 
the dot-induced mixing of the levels modifies 
the functional form of $G(u)$ in a non-trivial way.

\begin{figure}[p]
\centering

\includegraphics[width=6cm]{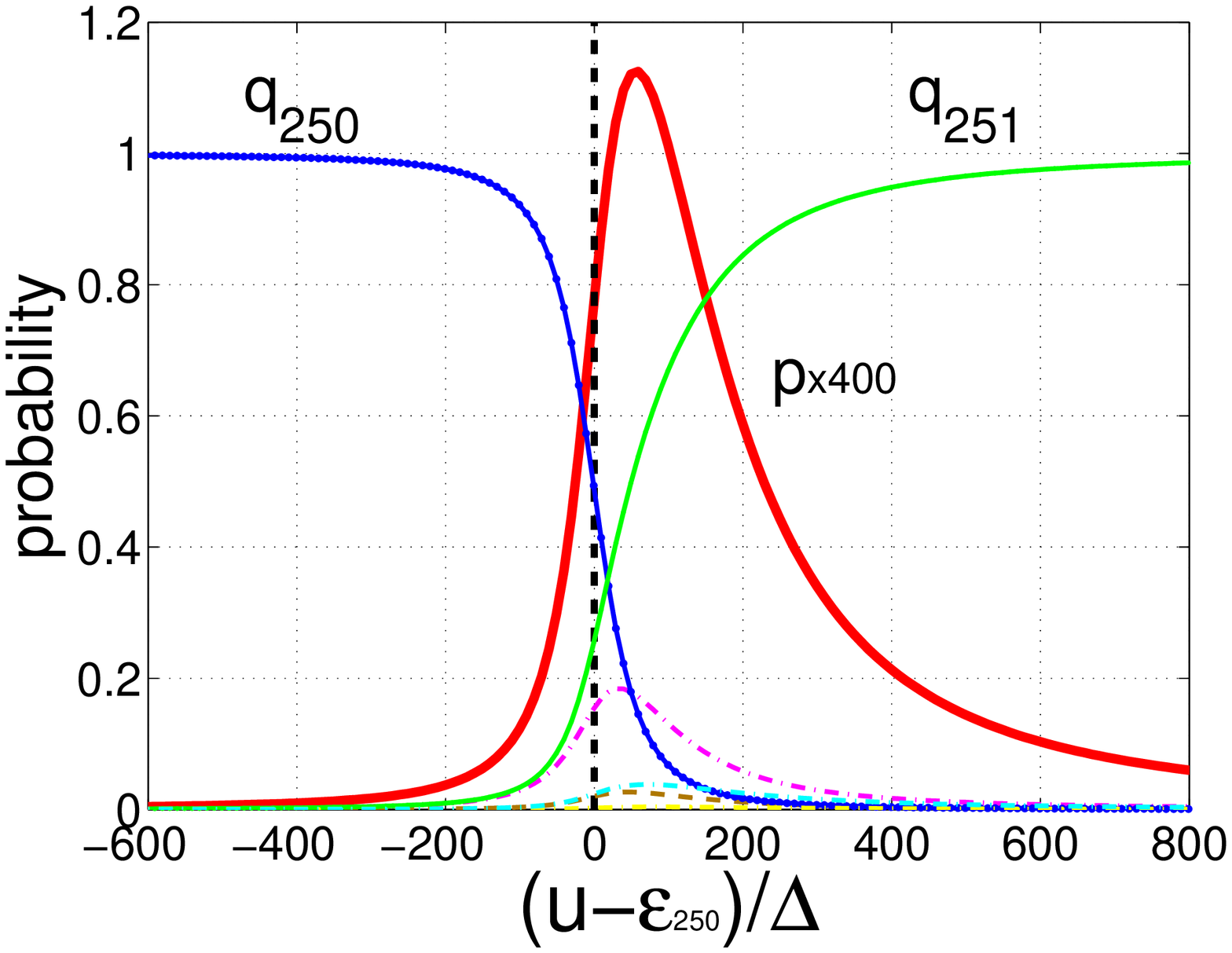}


\caption{
\rmrk{We consider an adiabatic sweep of an empty dot level
through a wire that is occupied by a single particle. 
The particle is initially placed at $n_0=250$ (an arbitrary level).}
The variation of the occupation probabilities is plotted as a function of $u$. 
The level spacing is ${\Delta=1}$, 
and the couplings are $C_a=6$ and $C_b=4$.
\rmrk{Accordingly the couplings to the even and to the odd levels 
are $c_{\pm}= (C_a \pm C_b)/\sqrt{2}$.} 
The red thick line is the dot occupation $p$, 
\rmrk{with vertical scale that is magnified ${\times 400}$.}
The other solid lines are $q_{250}$, and $q_{251}$.
The dashed lines from up to down 
are $q_{249}$ and $q_{253}$ and $q_{247}$ and $q_{252}$.
}
\label{fP}

\ \\ \ \\ 

\includegraphics[width=60mm]{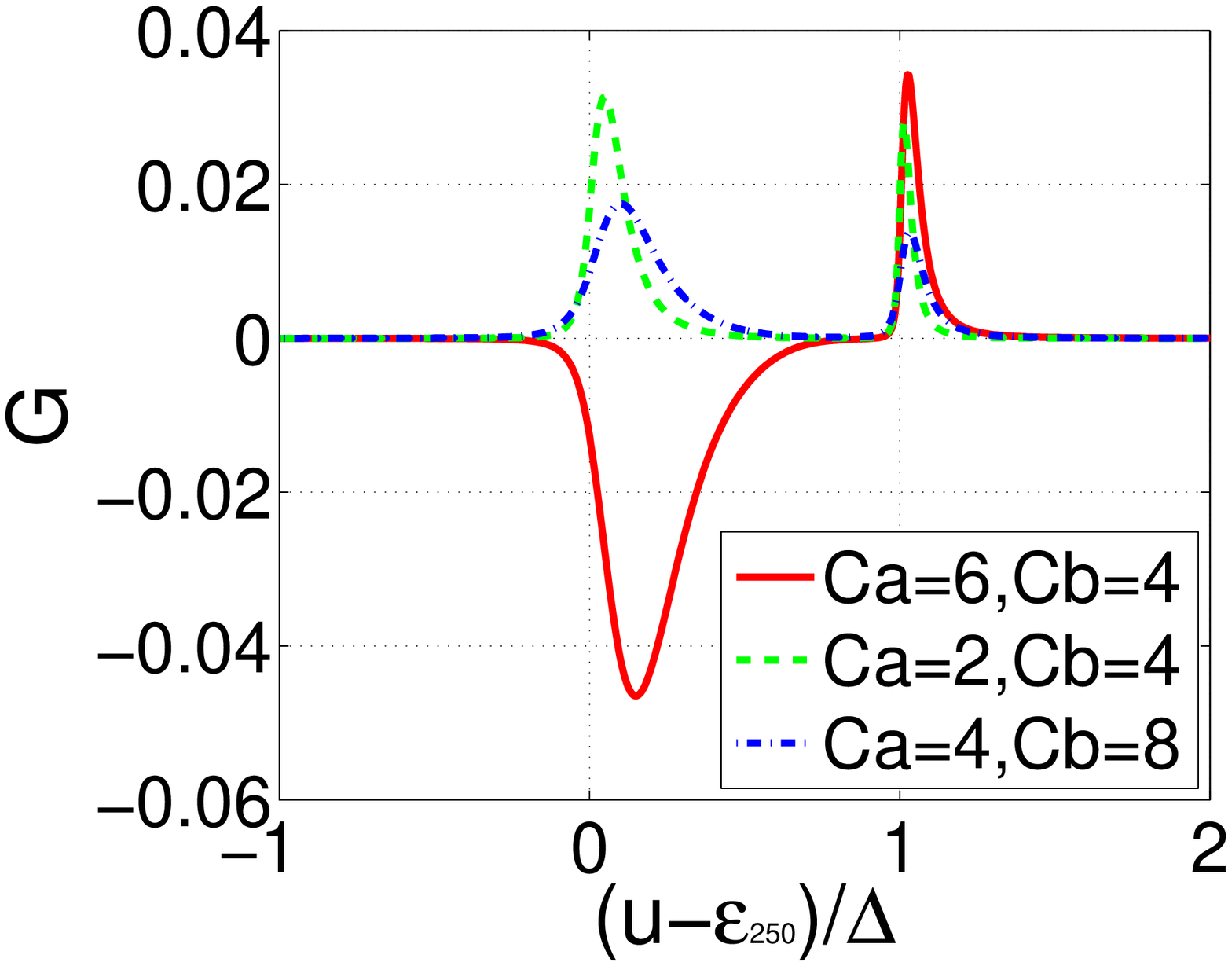} 

\includegraphics[width=60mm]{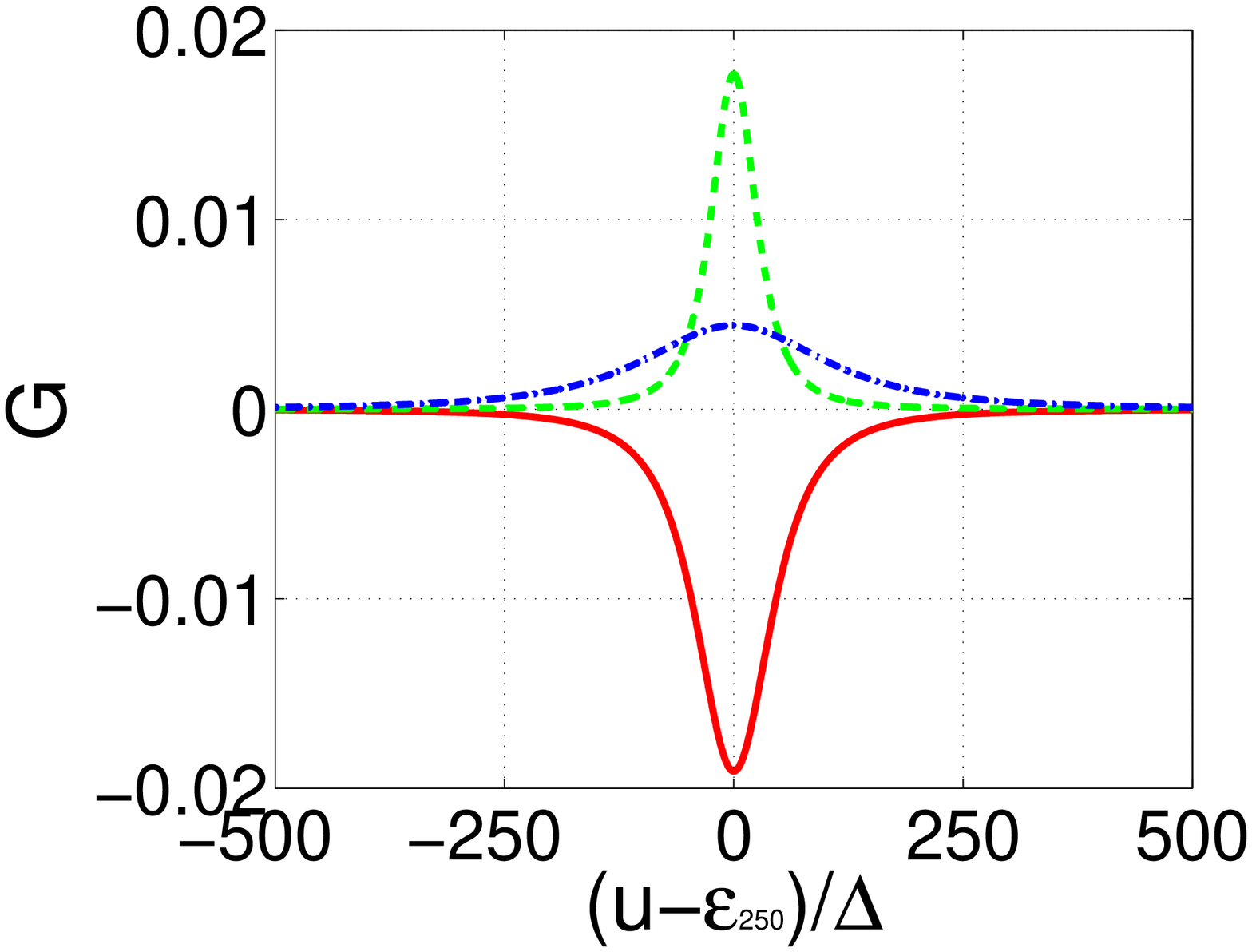}

\caption{Flow of the current from the dot
to the wire through the $C_a$ bond, 
for the same scenario as in \Fig{fP}. 
The parameters are indicted in the legend.
The raw calculation is done using \Eq{e9} with \Eq{e100}.
In the upper panel ${\Delta =200}$,
hence a two-level approximation \rmrk{for each crossing} is satisfactory.
In the lower panel ${\Delta = 1}$,
hence the explicit result \Eq{e10} can be optionally 
used in order to describe the multi-level crossing.
}
\label{fG}

\end{figure}

\rmrk{We now turn to discuss what happens with $N\gg1$ dot-wire system.
In \Fig{fP} we show how the occupations of the levels 
change as $u$ is swept during an adiabatic induction process.
Initially only level $n_0=250$ is occupied, while at the end 
of the sweep the probability is fully transferred to $n=251$.
The figure assumes ${c_n \gg \Delta}$, and therefore 
during the process many other levels are occupied. 
This is what we call quasi-continuum case.
In the other extreme of having  ${c_n \ll \Delta}$,
only 3 levels participate in the scenario: the dot level 
and the network levels ${n=250,251}$.    
In the latter case there are two distinct crossings, 
each can be described as a two-level crossing, 
with current dependence that is shown in the upper panel of \Fig{fG}.   
In contrast to that, in the quasi continuum case, 
individual crossings with the network levels cannot be resolved. 
Rather we see in the lower panel of \Fig{fG} 
that there is a single wide collective peak 
in the current that extends over an energy range 
that contains many network levels. It is the purpose 
of the next section to get an analytical understanding    
of this multi-level mixing, and to obtain an explicit 
result for the current dependence.}


\section{Adiabatic mixing in quasi continuum}

We turn to the detailed analysis of adiabatic mixing in 
the dot-wire system. The first step is to get an expression 
for $g(E)$ of \Eq{e40}. 
With $c_n=c_{\pm}$ the sum over the levels 
splits into two partial sums, 
over the odd and over the even levels. 
Consequently after summation we get two terms: 
\beq 
g(E) = 
\left(\frac{\pi}{2\Delta}\right) 
\left[
c_{-}^2 \cot\left(\pi\frac{E}{2\Delta}\right)  
-c_{+}^2 \tan\left(\pi\frac{E}{2\Delta}\right)  
\right]
\eeq
The secular equation ${g(E)=E-u}$ becomes a quadratic 
equation for $\cot()$, and can be solved explicitly:
\beq 
\cot
\left(\pi\frac{E}{2\Delta}\right) 
= 
\frac{\Delta}{\pi c_{-}^2} 
\left[
(E{-}u) \pm \sqrt{ (E{-}u)^2+ \left(\frac{\pi c_{+} c_{-}}{\Delta} \right)^2 }
\right]  
\eeq
where the $\pm$ refers to the parity that is alternating 
for subsequent levels. 
Then it is straightforward to get an explicit 
expression for the dot occupation probability $p(u)$
via \Eq{e100}, and for the level occupations $q_n(u)$ via \Eq{e102}.
The expressions are quite lengthy but can be simplified 
in the regime of interest as described below. 

Of interest is the case of a quasi-continuum,  
meaning that the couplings $c_n$ are larger compared with $\Delta$, 
hence a two level approximation is out of the question, 
while a Wigner-type approximation is most appropriate.   
For this purpose we find it useful to define parameters 
that describe the effective coupling of the dot 
to the quasi-continuum, and its asymmetry:
\beq
C_{\text{eff}} &\equiv& \ \ \frac{\pi}{2} \frac{c_{+}c_{-}}{\Delta}
\\ \label{e17}
\Gamma &\equiv& \pi \frac{c_{+}^2+c_{-}^2}{\Delta}
\\
\sin(\theta) &\equiv& \frac{c_{+}^2 -c_{-}^2}{c_{+}^2 + c_{-}^2} 
\eeq
Here and below we assume without loss of generality 
that the particle starts in an even-parity level.  
Using these notations we get after some algebra an approximation 
that should be valid in the quasi-continuum case:
\beq
p(u) \ \ \approx \ \ \Delta \cdot \mathrm{L}\left[u-E; \ \Gamma, \theta \right]
\eeq
The distorted Lorentzian $\mathrm{L}\left[x;\Gamma; \theta \right]$ is 
\beq 
\frac{1}{\pi}
\left[
1 
+
\frac{\sin\theta \ x} 
{\sqrt{x^2+\cos^2\theta \ (\Gamma/2)^2}}  
\right]^{-1}
\frac{\cos^2\theta \ (\Gamma/2)}
{x^2 + \cos^2\theta \ (\Gamma/2)^2} 
\eeq
In the expression above $E$ is the energy in which the 
particle has been prepared. In the regime of interest, 
where the levels are treated as quasi-continuum, 
this energy can be regarded as a constant. 
Some further straightforward algebra leads to
\be{9}
G(u) &=&
C_a 
\frac{\partial}{\partial u}
\left[ 
p \sum_n \frac{c_n^* \langle n|a \rangle }{(E-\epsilon_n)^2}  
\right]
\\ 
&=&
C_a \frac{\partial}{\partial u}
\left[ 
\frac{\frac{c_{-}}{\sin^2\left(\pi\frac{E}{2\Delta}\right)}
+\frac{c_{+}}{\cos^2\left(\pi\frac{E}{2\Delta}\right)}}
{\left(\frac{2\Delta}{\pi}\right)^2  
+\frac{c_{-}^2}{\sin^2\left(\pi\frac{E}{2\Delta}\right)}
+\frac{c_{+}^2}{\cos^2\left(\pi\frac{E}{2\Delta}\right)}}
\right]
\\ 
&\approx&
\frac{\partial}{\partial u} C_a
\left[ 
\frac{c_{+} + c_{-} \cot^2\left(\pi\frac{E}{2\Delta}\right)}
{c_{+}^2  + c_{-}^2 \cot^2\left(\pi\frac{E}{2\Delta}\right)} 
\right] 
\\ \label{e10}
&=&
(\lambda_{-}-\lambda_{+})\frac{2C_{\text{eff}}^2}{\left(4C_{\text{eff}}^2 + (u-E)^2\right)^{3/2}}
\eeq
Disregarding the splitting-ratio factor, this expression 
has surprisingly the same functional form as that 
of crossing a single level ($N=1$), see e.g. \Cite{pmr}, 
but with an effective coupling constant $C_{\text{eff}}$ 
that reflects the density of states.  

The functions $p(u)$ and $G(u)$ are plotted in \Fig{fP} and in \Fig{fG}.
In the latter we contrast with the ${c_n\ll \Delta}$ case,  
for which the dynamics can be regarded as a sequence 
of two $N=1$ crossings.


\section{Adiabatic and non-adiabatic regimes}

The results for the integrated current give another wrong impression: 
it looks as if we are dealing with two regimes: either the process 
is adiabatic or non-adiabatic. A more careful inspection reveals 
that depending on $\dot{u}$ we have 3~regimes: Adiabatic, Slow and Fast.  
For star geometry with comb-like quasi continuum of levels, 
the Slow regime is defined by the condition 
\be{7}
c^2 \ < \ \dot{u} \ < \ \Gamma^2, 
\ \ \ \ \ \ \Gamma\equiv 2\pi\frac{c^2}{\Delta}
\eeq 
For simplicity we assume here comb-like quasi continuum 
with identical couplings ${c_n=c}$.
The left inequality in \Eq{e7} means that the adiabatic 
condition is violated, while the right inequality 
implies that a first-order perturbative approximation is violated as well.
The identification of this intermediate Slow regime parallels 
the notion of Wigner or FGR or Kubo regime in past studies 
of time-dependent dynamics \Cite{pmc}.

Some illustrations for energy spreading are presented in \Fig{fS}.
If ${c<\Delta}$ the transport of probability from the dot to the 
network levels would be described using a two-level approximation. 
But the illustration in the upper panel assumes ${c>\Delta}$, 
hence many levels are mixed within a parametric range $\Gamma$.  
The time during which this mixing takes place is $\Gamma/\dot{u}$.
In the opposite limit of Fast sweep, 
which we further discuss below, the decay time of the probability 
to the quasi-continuum is $1/\Gamma$.

\section{Non adiabatic spreading}

The calculation of $I$ in the non-adiabatic regime 
requires knowledge of $q_n(t)$. For star geometry this  
calculation is a variant of the Wigner decay problem, 
and hence can be solved analytically: instead of 
a {\em fixed} level that decays into a quasi-continuum 
we have a {\em moving} level. The usual textbook 
procedure is followed \Cite{decay} leading to the 
following set of equations 
\beq
\partial_t \Psi_0 &=& \Big[-i u(t) - (\Gamma/2)\Big]\Psi_0 
\\ 
\partial_t \Psi_n &=& -i \epsilon_n \Psi_n - i c_n \Psi_0
\eeq
With ${u(t)=\dot{u}t}$ one obtains the solution   
\be{112}
q_n(t) \ = \ \left| c_n \int_0^t d\tau \ \exp\left( i\epsilon_n \tau - i\frac{\dot{u}}{2}\tau^2  - \frac{\Gamma}{2}\tau \right) \right|^2 
\eeq 
By inspection one observes that going 
from the Slow to the Fast regime, 
the spreading line shape changes 
from Lorentzian-type to Fresnel-type, 
as illustrated in the lower panel of \Fig{fS}.

\section{Summary}

\rmrk{Molecular motors and pumps are of great interest 
in various fields of Physics and Biology. 
Conceptually the major theme concerns the possibility 
to induce a circulating motion, or a circulating 
current, by some driving protocol. We use the term {\em stirring} 
rather than {\em pumping} in order to emphasize that 
closed geometry is concerned (no reservoirs).}   
Considering (e.g.) the unidirectional rotation 
of a molecular rotor \Cite{st1}, it is possibly allowed 
to be satisfied with a stochastic picture \Cite{Saar}
that relates the currents, via a ``decomposition formula", 
to rates of change of occupation probabilities. 
Once we turn (e.g.) to the analysis of pericyclic
reactions \Cite{manz} this is no longer possible. 
In the latter case the method of calculating 
electronic quantum fluxes had assumed that they
can be deduced from the continuity equation. 
Such procedure is obviously not applicable for 
(say) a ring-shaped molecule: due to the multiple path 
geometry there is no obvious relation between 
currents and time variation of probabilities. 

\rmrk{Nevertheless, we have found using elementary considerations,} 
that it is possible to replace the traditional adiabatic transport formula \Eq{e1} 
by a simple expression \Eq{e6}, that holds both in adiabatic and non-adiabatic 
circumstances. It can be regarded as a generalized multi-path generalization  
of the  continuity equation. Hence the problem of calculating currents   
is reduced to that of calculating time-dependent probabilities~$q_n(t)$  
as in the \rmrk{above mentioned} stochastic formulation. 

Our result \Eq{e6} is quite general. We have demonstrated 
its use in the very simple case of ``ring geometry", 
but it can be applied to any network configuration, 
and for any $u(t)$ time dependence. In particular one can 
use it in order to analyze a multi-cycle stirring process.   
Furthermore, the application of \Eq{e6} to a many-body system 
of non-interacting particles follows trivially, 
with~$q_n(t)$ that represent the actual occupations of the levels. 
  
It is important to realize that the  ``splitting ratio" \Eq{e5}
unlike the stochastic ``partitioning ratio" is not bounded within $[0,1]$. 
This observation has implications on the calculation 
of ``counting statistics" and ``shot noise" \Cite{count1,count2,count3}.

We have emphasized aspects that go beyond 
the familiar two-level approximation phenomenology, 
related to the scrambling of the network levels 
during the sweep process. The dot-induced 
mixing is reflected in the time dependence 
of the currents, but not in $Q$.

\ \\
{\bf Acknowledgments.-- }
This research has been supported by the Israel Science Foundation (grant No.29/11).


\clearpage
\end{document}